\documentclass[11pt]{article}
\usepackage[margin=1in]{geometry}
\usepackage{cprotect}
\usepackage{color}
\usepackage[colorlinks=true,linkcolor=blue,citecolor=blue,urlcolor=blue]{hyperref}
\usepackage{amssymb,amsmath}
\usepackage{amsthm}

\usepackage{graphicx}
\usepackage{helvet}
\usepackage[font=footnotesize,labelfont=bf]{caption}
\usepackage{longtable}

\usepackage{multirow}

\usepackage[super]{natbib}

\usepackage{bm}
\newcommand\blfootnote[1]{%
  \begingroup
  \renewcommand\thefootnote{}\footnote{#1}%
  \addtocounter{footnote}{-1}%
  \endgroup
}

\usepackage{setspace}
\numberwithin{equation}{section}
\theoremstyle{plain}

\newcommand{\JJL}[1]{\textcolor{black}{#1}}
\newcommand{\ww}[1]{\textcolor{black}{#1}}
\newcommand{\wl}[1]{\textcolor{black}{#1}}

\title{Issues arising from benchmarking single-cell RNA sequencing imputation methods}

\author{%
Wei Vivian Li\,$^{1}$ \JJL{and} Jingyi Jessica Li\,$^{2,3,*}$
}

\date{\vspace{-5ex}}

\begin{document}

\maketitle

\blfootnote{\\
$^{1}$ Department of Biostatistics and Epidemiology, Rutgers, The State University of New Jersey, NJ 08854\\
$^{2}$ Department of Statistics, University of California, Los Angeles, CA 90095-1554\\
$^{3}$ Department of Human Genetics, University of California, Los Angeles, CA 90095-7088\\
$^{*}$ To whom correspondence should be addressed.
Email: jli@stat.ucla.edu
}

% \begin{abstract}
% Here we discuss several questionable points in Huang \emph{et al.}, \textit{Nature Methods} (2018) and report a reanalysis of the data in their Figure 2. We find that the original analysis in Huang \emph{et al.} was not a fair or reasonable evaluation of three imputation methods: SAVER, scImpute, and MAGIC, for single-cell RNA sequencing data. Our reanalysis results suggest a different conclusion from Huang \emph{et al.} about the performance of the three imputation methods on downstream cell clustering analysis.
% \end{abstract}

%\noindent To the Editor:

\noindent On June 25\ww{th}, 2018, Huang \emph{et al.} \citep{huang2018saver} published a computational method SAVER on \textit{Nature Methods} for imputing dropout gene expression levels \citep{pierson2015zifa} in single cell RNA sequencing (scRNA-seq) data. 
Huang \emph{et al.} performed a set of comprehensive benchmarking analyses, including comparison with the data from RNA fluorescence in situ hybridization, to demonstrate that SAVER outperformed two existing scRNA-seq imputation methods,  scImpute \citep{li2018accurate} and MAGIC \citep{van2018recovering}.
However, their computational analyses were based on semi-synthetic data that the authors had generated following the Poisson-Gamma model used in the SAVER method. 
We have therefore re-examined Huang \emph{et al.}'s study. We find that the semi-synthetic data have very different properties from those of real scRNA-seq data and that the cell clusters used for benchmarking are inconsistent with the cell types labeled by biologists. We show that a reanalysis based on real scRNA-seq data and grounded on biological knowledge of cell types leads to different results and conclusions from those of Huang \emph{et al}. 
\\

\begin{figure}[!tbhp]
\centering
\includegraphics[width = .98\textwidth]{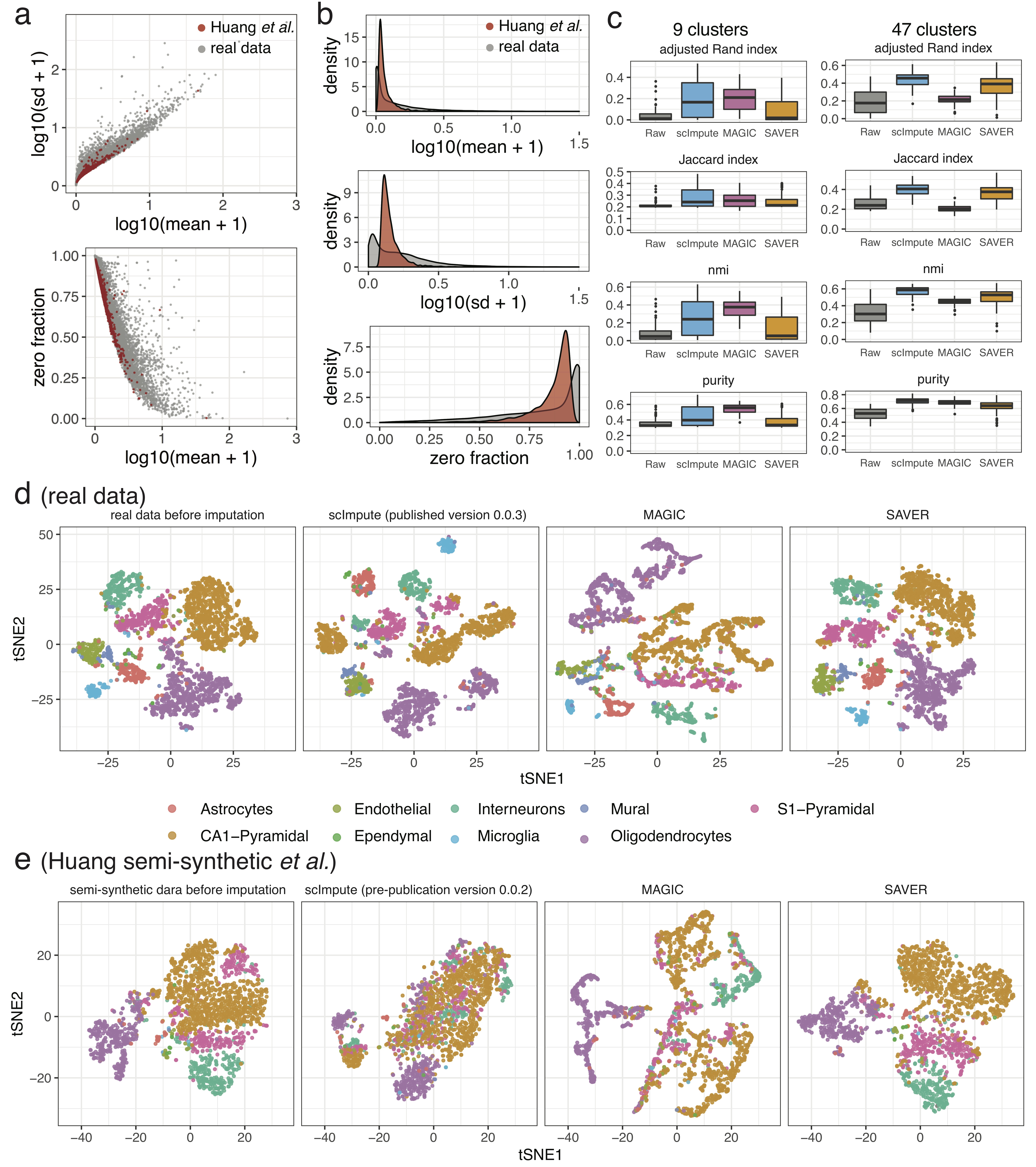}
\caption{Re-evaluation of Huang \emph{et al}. 
\textbf{a}: Comparison between the original scRNA-seq data from Zeisel \emph{et al.} for $19,912$ genes and $3,005$ cells \citep{zeisel2015cell} and the semi-synthetic data for $3,529$ genes and $1,800$ cells that Huang \emph{et al.} derived from the original data.
For genes with the same mean expression level, the semi-synthetic data exhibit a smaller standard deviation (sd) in gene expression (top panel) and a smaller fraction of zero expression (bottom panel) than the original data.
\textbf{b}: The distribution of gene expression mean, gene expression standard deviation, and per-gene fraction of zero count in both the original Zeisel \emph{et al.} data  and the semi-synthetic Huang \emph{et al.} data. 
\textbf{c}: Four evaluation measures (adjusted Rand index, Jaccard index, normalized mutual information, and purity) of the clustering results (using $K=9\ \text{and}\ 47$) on the Zeisel \emph{et al.} data and the three imputed datasets. Bootstrapping of cells were performed $100$ times to obtain the boxplots.
\textbf{d}: Two-dimensional tSNE representation of the original Zeisel \emph{et al.} data and the imputed data by scImpute (v0.0.3), MAGIC (v1.0.0), and SAVER (v1.0.0). The cells are colored based on the nine cell types from Zeisel \emph{et al}.
\textbf{e}: \wl{Two-dimensional tSNE representation of the semi-synthetic Huang \emph{et al.} data and the imputed data by scImpute (v0.0.2), MAGIC (v1.0.0), and SAVER (v1.0.0). The four datasets were generated in Huang \emph{et al.}'s analysis. The cells are colored based on the nine cell types in Zeisel \emph{et al}.}
\label{fig:zeisel}}
\end{figure}

\noindent To compare SAVER, scImpute, and MAGIC, Huang \emph{et al.} used four semi-synthetic datasets simulated based on the statistical assumptions of SAVER.
In detail, Huang \emph{et al.}  collected four real scRNA-seq datasets from public repositories, and they selected high-quality cells and highly-expressed genes from each dataset to make a \emph{reference dataset}. From these they created the semi-synthetic datasets by simulating gene expression levels from a Poisson-Gamma model that is used within the SAVER method. 
Specifically, they estimated a parameter $\lambda_{cg}$, i.e., the true expression level of gene $g$ in cell $c$, by the observed expression level of gene $g$ in cell $c$ in each reference data, and they denoted the parameter estimate by $\hat\lambda_{cg}$. 
They also randomly sampled a coefficient $\tau_c$ for every cell $c$ from an arbitrary Gamma distribution. Then they simulated $X_{cg}$, the semi-synthetic expression level of gene $g$ in cell $c$, by randomly sampling a value from $\text{Poisson}(\tau_c \hat\lambda_{cg})$.
\\

\noindent We find, however, that  Huang \emph{et al.}'s four semi-synthetic datasets underrepresent the proportions of zero \JJL{gene expression levels} and the heterogeneity of gene expression levels across various cells, compared with the four original real datasets (Fig. \ref{fig:zeisel}a). Moreover, the distributions of gene expression mean and standard deviation, as well as per-gene fraction of zero read count  in the semi-synthetic data demonstrate substantial differences compared with the real data (Fig. \ref{fig:zeisel}b).
Also, the average correlation between a  given gene's expression levels in the real data and those in the synthetic data is poor. The average coefficient of determination is only $R^2 = 0.14$, which means that on average the semi-synthetic data can  only explain $14\%$ of each gene's variation in the real data. Thus, these  semi-synthetic datasets may have significantly different properties from those of real data, and computational results based on these semi-synthetic data should have been interpreted in a more cautious way.

\begin{table}[!tbhp]
\caption{\label{tab}
The contingency table of actual cell types defined by Zeisel \emph{el al.} using biological marker genes versus the cell types used by Huang \emph{el al.} to analyze their semi-synthetic data. 
Huang \emph{el al.} generated a semi-synthetic dataset of $3,529$ genes and $1,799$ cells from the Zeisel \emph{el al.} dataset of $19,912$ genes in $3,005$ cells. They compared different imputation methods using seven cluster labels ($0, 1,\dots, 6$) that were defined using the Seurat algorithm \citep{satija2015spatial}. Each column in the table lists the actual cell type composition of cells grouped into one of the clusters defined by Huang \emph{el al}. Actual cell types were identified based on the presence of cell-type parker genes. 
}
\centering
\includegraphics[width = .6\textwidth]{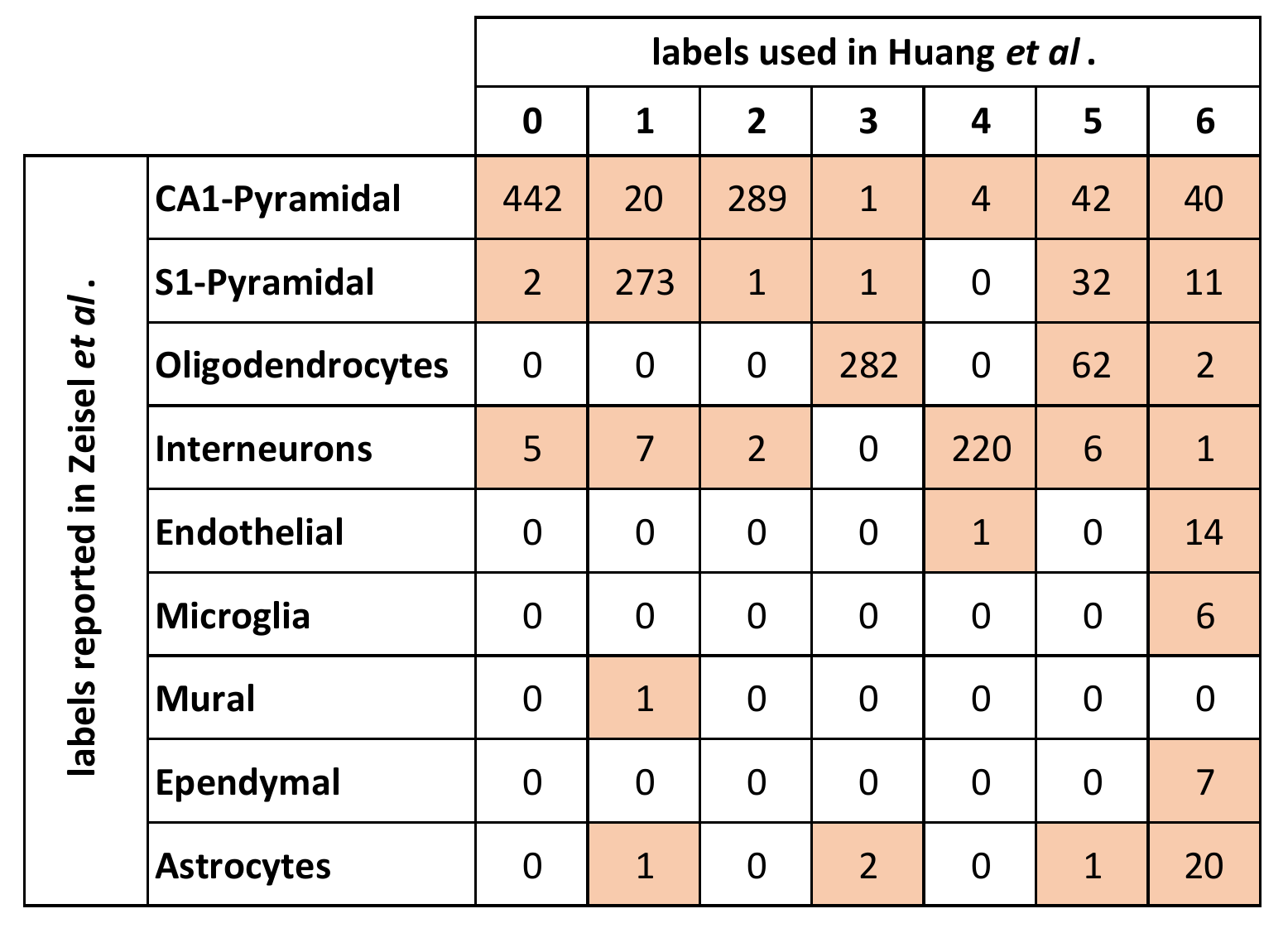}
\end{table}

\noindent To benchmark the SAVER method, Huang \emph{et al.} evaluated the assignment of cells to a defined set of clusters for the imputed data and, separately, for the semi-synthetic data. For this analysis, they used seven cell cluster labels that had been defined by the computational method Seurat \citep{satija2015spatial} based on the reference datasets. Crucially, though, no prior biological knowledge was used to define the seven clusters, and we have found that the clusters do not correspond well to the cell-type marker genes. For example, the scRNA-seq data for $19,912$ genes in $3,005$ cells, published by Zeisel \emph{et al.}, were shown to have nine major cell types and $47$ subtypes using known marker genes and cellular functions \citep{zeisel2015cell}. The reference dataset down-sampled by Huang \emph{et al.} from Zeisel \emph{et al.}'s data contained values for only $3,529$ genes in $1,799$ cells, and the seven cluster labels assigned to these cells do not agree with Zeisel \emph{et al.}'s cell types (Table \ref{tab}). In addition, Huang \emph{et al.} used the Seurat software to benchmark the clustering of cells based on the imputed datasets or the reference dataset. It is unclear what results would be obtained using alternative clustering methods other than Seurat. 
\\

\noindent For the above reasons, we were concerned that the results in Huang \emph{et al.} may not reflect the actual performance of imputation methods, when applied to real data to identify biologically relevant cell types. Therefore, we re-evaluated the performance of SAVER, scImpute, and MAGIC, also from the perspective of cell clustering, by directly using the original data from Zeisel \emph{et al}. 
We asked if the cell clusters found resemble the nine major cell types and the $47$ subtypes reported in Zeisel \emph{et al}. To answer this question, we performed hierarchical clustering with $K = 9$ and $K = 47$ on both the original and the imputed data, based on their first ten principal components. Please note that the original data represent the whole Zeisel \emph{et al.} dataset, not the reference dataset selected by Huang \emph{et al.} Evaluation of the clustering results indicates that data imputed by scImpute leads to comparable or higher adjusted Rand index \citep{hubert1985comparing}, Jaccard index \citep{milligan1986study}, normalized mutual information (nmi) \citep{witten2016data}, and purity (Figure \ref{fig:zeisel}c), compared with the data imputed by SAVER for both $K = 9$ and $K = 47$. This result contradicts Huang \emph{et al.}'s conclusion that ``SAVER achieved a higher Jaccard index than that observed for all datasets, whereas MAGIC and scImpute had a consistently lower Jaccard index''. We also visualized the gene expression data before and after imputation by each method using the t-distributed stochastic neighbor embedding (tSNE). Our tSNE visualization suggests that based on the real data, the all the three imputation methods lead to clear separation patterns for nine biologically defined cell types (Fig. \ref{fig:zeisel}d). However, Huang \emph{et al.}'s semi-synthetic data show a highly different visualization (Fig. \ref{fig:zeisel}e).
\\

\noindent Our analysis was carried out using the Ubuntu 14.04.5 system and 2 CPUs of Intel(R) Xeon(R) CPU E5-2687W v4 @ 3.00GHz. The running time of SAVER (version 1.0.0, \url{https://github.com/mohuangx/SAVER}), scImpute (version 0.0.3, \url{https://github.com/Vivianstats/scImpute}), and MAGIC (version 1.0.0, \url{https://github.com/KrishnaswamyLab/MAGIC}) were $2430.85$s, $1519.78$s, and $21.14$s, respectively. We ran scImpute using $35$ cores and default settings, with \verb|KCluster|=9. We ran MAGIC and SAVER with the default settings and $35$ cores.
In addition, we note that Huang \emph{et al.} used scImpute version 0.0.2, which was an archived version. The scImpute paper \citep{li2018accurate} improved the methodology and introduced scImpute version 0.0.3, which was released on October 22th, 2017. The scImpute package 0.0.3 includes an important step for the identification of cell sub-populations, which could significantly improve the accuracy and robustness compared to version 0.0.2.
Scripts for analysis in this article can be found at \url{https://github.com/Vivianstats/scImpute/tree/master/inst/comparison}.
\\

\noindent We appreciate the contribution of SAVER as a new Bayesian imputation method to borrow information across both genes and cells. 
Our results, however, suggest that the semi-synthetic data generated from the Poisson-Gamma mixture model in Huang \emph{et al.}  do not represent multiple key characteristics of real scRNA-seq data. This finding emphasizes the necessity of using real data in addition to synthetic data for reproducible research in the field of computational biology. Given that large-scale, error-free scRNA-seq data are not yet available for benchmarking, it remains critical to assess the performance of computational methods from perspectives that have biologically meaningful interpretations. As improved quality scRNA-seq data become available, we will be better equipped to perform comprehensive and fair comparisons of scRNA-seq computational methods. We suggest that all computational methods should make their assumptions and evaluation approaches clear and understandable to users, so users can fairly evaluate the biological relevance, advantages and drawbacks of each method before applying it to make scientific discoveries.

%\subsection*{Code availability}
%The codes used for the reanalysis are available at \url{https://github.com/Vivianstats/scImpute/tree/master/inst/comparison}.

\subsection*{Data availability}
The Zeisel \emph{et al.} data are available at the Gene Expression Omnibus (GEO) under accession code GSE60361.

\subsection*{Acknowledgement}
We thank Dr. Mark Biggin at Lawrence Berkeley National Laboratory for the discussions and suggestions, which helped us improve the manuscript.

% \vspace{ 1 em}
% \noindent\textbf{Author contributions}: 
% J.J.L designed the research. W.V.L conducted the analysis. W.V.L and J.J.L. discussed the results and contributed to the manuscript writing.

% \vspace{ 1 em}
% \noindent\textbf{Competing interests}: The authors declare no competing interests.

\clearpage
\bibliographystyle{unsrt}
\bibliography{scrna}

\end{document}